\documentclass[twocolumn,pra,aps,superscriptaddress, showpacs, amsmath,amssymb,10pt,longbibliography]{revtex4-2} 
\usepackage{amsmath} 
\usepackage{amssymb} 
\usepackage{amsfonts} 
\usepackage{bm} 
\usepackage{amssymb} 
\usepackage{graphicx} 
\usepackage{dcolumn} 
\usepackage{txfonts} 
\usepackage{wasysym} 
\usepackage{makeidx}
\usepackage{color}
\usepackage{mathtools}
\usepackage{threeparttable}
\usepackage[linkcolor=blue,anchorcolor=black,citecolor=blue,colorlinks=true]{hyperref}
\usepackage{breakurl}
\usepackage{float}
\usepackage{soul}
\usepackage{adjustbox}
\usepackage{autobreak} 
\usepackage{cleveref}
\usepackage{booktabs}  
\usepackage{array}     
\usepackage{siunitx}   

\begin{document}
\title{Impact of Heavy Noble Gases on the Magnetic Resonance Linewidth of Alkali-Metal Atoms: A Theoretical Study}
\author{Feng Tang}
\affiliation{Beijing Computational Science Research Center, Beijing 100193, PR China}%
\author{Kezheng Yan}
\affiliation{Beijing Computational Science Research Center, Beijing 100193, PR China}%
\author{Nan Zhao}
\email{nzhao@csrc.ac.cn}
\affiliation{Beijing Computational Science Research Center, Beijing 100193, PR China}%
\date{\today}

\begin{abstract}

Nuclear magnetic resonance gyroscopes (NMRGs) employ noble-gas nuclear spins as inertial sensors and alkali-metal atoms as in-situ magnetometers. Heavy noble gases, particularly xenon, are widely used due to their large nuclear spin and strong spin-exchange coupling with alkali-metal atoms. However, their presence introduces additional collisional mechanisms that affect the alkali-metal magnetic resonance linewidth, thereby influencing magnetometer sensitivity and overall gyro performance.
In this work, we develop a theoretical framework based on the density matrix formalism and master equation approach to quantitatively study how xenon-induced two-body and three-body interactions modify the linewidth of alkali-metal atoms under realistic NMRG conditions. Our analysis reveals that Xe atoms primarily broaden the linewidth via binary spin-destruction collisions and van der Waals (vdW)-mediated F-damping processes, while the effect of Xe nuclear polarization is negligible at the $\sim 1\%$ level.
We further demonstrate that nitrogen buffer gas plays a dual role: it directly contributes to alkali-metal spin relaxation through binary collisions and indirectly modulates vdW collision rates by altering molecular lifetimes. The interplay between these processes leads to an optimal nitrogen density that minimizes the linewidth. Additionally, we identify a temperature threshold above which light-narrowing emerges, with this threshold increasing alongside Xe density. These findings provide theoretical insight for optimizing spin relaxation control in alkali-metal magnetometers and improving NMRG performance.
\end{abstract}

\maketitle

\section{Introduction}

Nuclear magnetic resonance gyroscopes have emerged as a promising class of high-precision inertial sensors that operate without reliance on external references. Their compact size, low power consumption, and excellent long-term frequency stability make them well-suited for navigation applications in aerospace, marine, and terrestrial environments \cite{Donley2010,Kitching2011,Fang2012,Meyer2014,Donley2013}. By exploiting nuclear magnetic resonance (NMR) techniques, NMRGs achieve exceptional sensitivity to rotational motion over extended time scales.

The working principle of NMRGs is based on detecting the precession of polarized noble-gas nuclear spins. The system comprises two key atomic species: alkali-metal atoms (such as rubidium), which act as in-situ magnetometers, and noble-gas atoms (such as xenon), which serve as the inertial sensing medium \cite{Walker2016,Eklund2008}. Spin-exchange optical pumping is used to transfer angular momentum from optically pumped alkali-metal electrons to the noble-gas nuclei, thereby polarizing the latter. Once polarized, the noble-gas nuclear spins precess in the presence of a transverse driving field, generating a rotating magnetic field. This field is detected by the alkali-metal magnetometer. When the cell undergoes mechanical rotation, the nuclear precession frequency shifts due to the gyroscopic effect, enabling rotation sensing with high precision.
As a key component of NMRGs, the sensitivity of the alkali-metal magnetometer directly impacts the overall performance of the system \cite{Gao2024}. This sensitivity is fundamentally limited by the magnetic resonance linewidth of the alkali-metal atoms \cite{Budker2007,Budker2013}, which is governed by various spin-relaxation processes. Among the dominant mechanisms are two-body collisions, including: (i) spin-exchange (SE) and spin-destruction (SD) collisions between alkali-metal atoms (e.g., Rb–Rb), (ii) SE and SD collisions between alkali-metal and noble-gas atoms (e.g., Rb–Xe), and (iii) SD collisions between alkali-metal atoms and buffer or quenching gases such as nitrogen (Rb–$\rm{N}_2$) \cite{Happer1977,Appelt1998,Seltzer2008}. These collisions cause decoherence of the alkali-metal spins, directly contributing to the observed resonance linewidth and thus setting fundamental limits on magnetometric sensitivity.

In the case of heavy noble gases, particularly xenon, the formation of alkali-metal–noble-gas van der Waals (vdW) molecules introduces additional relaxation channels for the alkali-metal spins \cite{Appelt1998,Bouchiat1972,Zeng1985,Happer1984}. During the lifetime of these transient molecules, spin-rotation and nuclear-electron spin-exchange interactions contribute to both S-damping and F-damping of the alkali-metal spins.  The formation rate and lifetime of vdW molecules are strongly dependent on the partial pressure of nitrogen, which acts as the third-body collision partner, and thus $\mathrm{N}_2$ plays a dual role in determining the overall linewidth.

In this work, we present a theoretical study of the alkali-metal linewidth in NMRGs, with particular emphasis on the influence of xenon on rubidium atoms. We develop a master-equation framework that incorporates spin-exchange and spin-destruction collisions, as well as three-body interactions mediated by molecule formation. By solving the equations under the rotating wave and weak-driving approximations, we find that Xe broadens the Rb resonance primarily through binary SD collisions and vdW-induced F-damping. A competition between Rb–$\mathrm{N}_2$ spin-destruction and vdW processes gives rise to an optimal nitrogen pressure that minimizes the linewidth. The combined SD effect from Rb--Xe binary and vdW collisions suppresses the light-narrowing phenomenon, which only becomes observable once the Rb--Rb spin-exchange rate exceeds a critical value, typically achieved above a characteristic temperature threshold. These results provide a quantitative understanding of collisional relaxation of  rubidium atoms in NMRGs and offer guidance for optimizing alkali-metal magnetometry in such systems.

\section{Theoretical treatment}
\subsection{Master equation in the Liouville space}
The system under investigation consists of a gas cell containing alkali-metal Rb, nitrogen ($\mathrm{N}_2$), and noble gas $^{129}\mathrm{Xe}$. The Rb atoms are optically pumped by a $\sigma_+$ laser beam, resonant with the $D_1$ line, propagating along the $z$-axis. Additionally, the Rb atoms are subjected to a static magnetic field $B_0$ along the z-axis and a radio-frequency (RF) magnetic field along the x-axis. The evolution of the ground-state Rb atoms in the high-pressure broadening limit is governed by \cite{Happer2010,Appelt1998}
\begin{align}\label{MasterEq_H}
\frac{d}{dt}\rho &= -i[H,\rho] +\Gamma_{\mathrm{SD}}(\varphi-\rho)+ \Gamma_{\mathrm{SE}}^{\mathrm{RbRb}}[\varphi(1+ 4 \langle \mathbf{S} \rangle \cdot \mathbf{S})-\rho] \nonumber \\
&+ \Gamma_{\mathrm{SE}}^{\mathrm{RbXe}}\left[\varphi(1+4 \langle \mathbf{K}\rangle\cdot \mathbf{S})-\rho \right]+ R_{\mathrm{op}} \left[\varphi(1+2 \mathbf{s}\cdot \mathbf{S})-\rho \right]  \nonumber \\
&+ \frac{2 \phi_{\gamma}^2}{3T_{\mathrm{vdwA}}}\left[ f_{s} (\varphi-\rho) + 
\frac{f_{F}} {[I]^2}(\mathbf{F}\cdot \rho \mathbf{F}-\frac{1}{2}\{{\mathbf{F}\cdot\mathbf{F},\rho\}})
\right]  \nonumber \\
&+ \frac{\phi_{\alpha}^2}{2T_{\mathrm{vdwA}}}\bigg[
f_{s} \left(\varphi(1+4 \langle \mathbf{K} \rangle \cdot \mathbf{S})-\rho \right) + \frac{f_F}{[I]^2} \times \nonumber\\ 
&
 \big[  \mathbf{F}\cdot\rho \mathbf{F}-\frac{1}{2}\{{\mathbf{F}\cdot\mathbf{F},\rho}\}          
+ (\{\mathbf{F},\rho\}-2i  \mathbf{F}\times \rho \mathbf{F})\cdot \langle \mathbf{K} \rangle \big]
\bigg].
\end{align}

The Hamiltonian governing the coherent evolution of the Rb atoms is given by 
\begin{align}
H = H_0 + H_{\mathrm{drv}},
\end{align}
where $H_0$ describes the interaction with the static magnetic field 
\begin{align}
H_0 = \Omega_{0}S_{z} + \Omega_{I} I_{z} + \Omega_{\mathrm{hf}} \mathbf{I} \cdot \mathbf{S},
\end{align}
and $H_{\rm{drv}}$ represents the interaction with the RF field 
\begin{align}
H_{\mathrm{drv}} = \Omega_{R} S_x \cos(\omega t).
\end{align}
The Hamiltonian $H_0$ consists of the Zeeman interaction $\Omega_0 S_z + \Omega_I I_z$ and the hyperfine interaction $\Omega_{\mathrm{hf}} \mathbf{I}\cdot \mathbf{S}$, where $\Omega_0 = |\gamma_e|B_0$ stands for the Larmor frequency of the electron spin, $\Omega_I = \gamma_I B_0$ represents the Larmor frequency of the nuclear spin and $\Omega_{\mathrm{hf}}$ denotes the hyperfine interaction constant. Throughout this work, we assume that the Zeeman interaction is significantly weaker than the hyperfine interaction, i.e., $\Omega_0 \ll \Omega_{\mathrm{hf}}$.
The RF field $\Omega_R \cos(\omega t)$ induces Zeeman transitions within a given hyperfine multiplet, where $\Omega_R$  is the Rabi frequency  and $\omega$ is the  RF driving frequency.

The incoherent evolution of the density matrix $\rho$ results from various collisions, including spin-exchange (SE) collisions, spin-destruction (SD) collisions, and three-body van der Waals (vdW) molecular processes. The second term in Eq.~\eqref{MasterEq_H} accounts for SD collisions between alkali-metal atoms and other atoms or molecules, with the total SD rate denoted as $\Gamma_{\mathrm{SD}}$. The third term in Eq.~\eqref{MasterEq_H} represents the SE collisions between Rb atoms, occurring at a rate of $\Gamma_{\mathrm{SE}}^{\mathrm{RbRb}}$, where $\langle \mathbf{S}\rangle = \mathrm{Tr} (\mathbf{S}\rho)$ is the electronic-spin expectation value. The fourth term describes the SE collisions between Rb and $^{129}\mathrm{Xe}$, with a corresponding collision rate $\Gamma_{\mathrm{SE}}^{\mathrm{RbXe}}$, where $\langle \mathbf{K} \rangle$ denotes the nuclear-spin expectation value of $^{129}\mathrm{Xe}$.

Beyond binary collisions, the evolution of $\rho$ is also influenced by vdW molecular processes, as characterized by the last two terms in  Eq.~\eqref{MasterEq_H}. The second-to-last term of Eq.~\eqref{MasterEq_H} originates from the spin-rotation interaction, $\gamma \mathbf{N}\cdot \mathrm{S}$, where $\mathbf{N}$ is the relative angular momentum of the colliding pair of atoms and $\gamma$ is the coupling coefficient \cite{Appelt1998}. As demonstrated by Appelt \cite{Appelt1998}, a fraction $f_s$ of vdW molecules exhibits "very short" lifetimes, while the remaining fraction $f_F$ has "short" lifetimes. It is noticed that a fraction of $f_F$ vdW molecules  undergoes transitions with $\Delta F=0$, where $F$ represents the spin quantum number of the total angular momentum $\mathbf{F}=\mathbf{I}+\mathbf{S}$.

The final term in Eq.~\eqref{MasterEq_H}  arises from the nuclear-electron SE interaction $\alpha \mathbf{S}\cdot \mathbf{K}$, with couping coefficient $\alpha$. The quantities $\phi_\alpha$ and $\phi_{\gamma}$ represent the mean precession phase angles induced by the nuclear-electron SE and spin-rotation interactions, respectively. The formation rate of vdW molecules per Rb atom is denoted by $1/T_{\mathrm{vdWA}}$ (see Appendix \ref{vdWProcess} for further details). Finally, the symbol $[I]$ is defined as $[I]=2I+1$, where $I$ is the quantum number of the nuclear spin.

The $D_1$ line pumping is described by the fifth term of Eq.~\eqref{MasterEq_H}, where $R_{\mathrm{op}}$ denotes the pumping rate and $\mathbf{s}$ signifies the mean photon spin. The mean photon spin $\mathbf{s}$ is determined by the polarization of the pumping light \cite{Happer1972,Happer2010}:
\begin{align}
\mathbf{s} = i\mathbf{e} \times \mathbf{e}^{*},
\end{align}
where $\mathbf{e}$ is the unit polarization vector of the pumping light. In this work, we consider the longitudinal pumping with $\sigma_+$-polarized light, corresponding to $\mathbf{s}=\hat{z}$. The state 
\begin{align}
\varphi = \frac{1}{4}\rho  + \mathbf{S}\cdot \rho \mathbf{S}
\end{align}
represents a purely nuclear spin operator of the alkali-metal atom, devoid of electronic-spin polarization \cite{Happer2010,Happer1977,Appelt1998}.

In the context of Liouville space, the density matrix $\rho$ is mapped onto a state vector $|\rho)$. This state vector $|\rho)$ can be expressed in the basis
\begin{equation}
\vert m_F; m'_{F'}) = \vert F, m\rangle \langle F', m'\vert 
\label{eq:basis} 
\end{equation}
 as
\begin{equation} 
|\rho) = \sum_{m,m',F,F'} |m_F;m'_{F'})(m_F;m'_{F'}|\rho ).
\end{equation}
where $\vert F,m \rangle$ is the eigenvector of total angular momentum $F_z$, and $(m_F;m'_{F'}|\rho )=\mathrm{Tr}[(\vert F,m\rangle \langle F' m'\vert)^{\dagger}\rho]$ is the inner product in Liouville space \cite{Appelt1998,Happer2010}. 

For alkali-metal atoms in the ground state, the total angular momentum quantum number $F$ takes the values $F = I + 1/2 \equiv a$ or $F = I - 1/2 \equiv b$, while the magnetic quantum number $m$ ranges from $-F$ to $F$. When considering state vectors restricted to the subspace of population and Zeeman coherences ($F=F'$), we define
\begin{equation} \vert F, \bar{m})_{\Delta m} = \vert m_F; m'_F), 
\label{eq:BasisZeeman} 
\end{equation} 
where $\bar{m}= (m+m')/2$ and $\Delta m = m-m'$ \cite{Appelt1998}.

Rewritting the master equation ~\eqref{MasterEq_H} in the Liouville space, we get
\begin{align}\label{MasterEqL}
\frac{d}{dt} |\rho) =& - i (\mathcal{H}_0^{\copyright}+ \mathcal{H}_{\mathrm{drv}}^{\copyright}  ) |\rho)
- \Gamma_{\rm{FD}} \mathcal{A}_{\rm{FD}} \vert \rho )
- \Gamma_{\rm{tot}} \mathcal{A}_{\text{SD}}|\rho)
 \nonumber\\
& + \left(\Gamma_{\text{SE}}^{\mathrm{RbRb}} ( \mathbf{S}\vert \rho ) +  \frac{1}{2}R_{\rm op}\hat{z}+ \Gamma_{\mathrm{SET}}^{\mathrm{RbXe}} \langle \mathbf{K} \rangle \right) \cdot \boldsymbol{\mathcal{A}}_{\text{SE}} |\rho) \nonumber\\
&+ \Gamma_{\mathrm{FE}}\langle \mathbf{K} \rangle \cdot \boldsymbol{\mathcal{A}}_{\mathrm{FE}} \vert \rho) \nonumber \\
&\equiv  -\mathcal{G} |\rho), 
\end{align}
where the relaxation parameters are
\begin{align}
\Gamma_{\mathrm{FE}}&=\frac{f_{F} }{2[I]^2} \frac{\phi_{\alpha}^{2}}{ T_{\mathrm{vdwA}}},\\
\Gamma_{\rm{FD}}&= \frac{f_{F}}{[I]^2} \left( \frac{2\phi_{\gamma}^2}{3T_{\rm{vdwA}}} + \frac{\phi_{\alpha}^2}{2T_{\rm{vdwA}}}   \right) ,\\
\Gamma_{\mathrm{SET}}^{\mathrm{RbXe}}&= \Gamma_{\mathrm{SE}}^{\mathrm{RbXe}}+ \Gamma_{\rm{SE,vdW}}^{\rm{RbXe}},\\
\Gamma_{\rm{tot}}& =  \Gamma_{\mathrm{SE}}^{\mathrm{RbRb}}+ R_{\mathrm{op}}+ \Gamma_{\mathrm{SET}}^{\mathrm{RbXe}} + \Gamma_{\rm{SD,vdW}}^{\rm{RbXe}} + \Gamma_{\mathrm{SD}},
\end{align}
and the electron-spin expectation value in Liouville space is given by $\langle \mathbf{S} \rangle = (\mathbf{S}\vert \rho )$. During the short-lived existence of Rb–Xe vdW molecules, nuclear-electron spin-exchange interactions lead to an effective additional spin-exchange rate given by
\begin{align} \Gamma_{\rm{SE,vdW}}^{\rm{RbXe}} & = \frac{f_{s}\phi_{\alpha}^2}{2T_{\mathrm{vdwA}}}, \end{align}
while both spin-rotation and spin-exchange interactions contribute to the total spin-destruction rate
\begin{align} \Gamma_{\rm{SD,vdW}}^{\rm{RbXe}} &= f_{s} \left( \frac{2 \phi_{\gamma}^2}{3T_{\mathrm{vdwA}}} + \frac{\phi_{\alpha}^2}{2T_{\mathrm{vdwA}}} \right). \end{align}

The Liouvillian superoperators defined in equation \eqref{MasterEqL} are \cite{Happer2010,Feng2025}
\begin{align}
\mathcal{A}_{\text{SD}} &= \frac{1}{2} \mathbf{S}^{\copyright} \cdot \mathbf{S}^{\copyright},\\
\mathcal{A}_{\text{FD}} &= \frac{1}{2}(\mathbf{F}^{\flat}\cdot \mathbf{F}^{\flat} +\mathbf{F}^{\sharp}\cdot \mathbf{F}^{\sharp})- \mathbf{F}^{\flat}\cdot \mathbf{F}^{\sharp}, \\
\boldsymbol{\mathcal{A}}_{\text{SE}} &= \mathbf{S}^{\flat} + \mathbf{S}^{\sharp} - 2 i \mathbf{S}^{\flat} \times \mathbf{S}^{\sharp},\\
\boldsymbol{\mathcal{A}}_{\text{FE}} &= \mathbf{F}^{\flat} + \mathbf{F}^{\sharp} - 2 i \mathbf{F}^{\flat} \times \mathbf{F}^{\sharp},
\end{align}
where $X^{\flat}$ and $X^{\sharp}$ denote the left- and right-translation superoperators \cite{Ernst1990}, and $X^{\copyright} = X^{\flat} - X^{\sharp}$ the commutator superoperator \cite{Happer2010}.

\subsection{Master equaiton under RWA and WDA}

We consider the case of a near-resonant RF driving field, where the driving frequency $\omega$ is close to the Larmor frequency $\Omega_0$, i.e., $\Omega_0 \approx \omega$. For typical NMRG applications, the static magnetic field $B_0$ is on the order of $10^4~\rm{nT}$, corresponding to a Larmor frequency of $\Omega_0 \sim 100~\rm{kHz}$ for $^{87}\rm{Rb}$. Outside the spin-exchange-relaxation-free (SERF) regime, where the dominant spin exchange rate $\Gamma_{\rm{SE}}^{\rm{RbRb}} \ll \Omega_0$, the rotating wave approximation (RWA) can be employed to simplify the master equation. 

Following the calculations in \cite{Feng2025}, we get the RWA equation:
\begin{equation} \label{eq:RWA_master_equation}
\frac{d}{dt} |\tilde{\rho}) = - \mathcal{G}_{\text{RWA}}|\tilde{\rho}), 
\end{equation}
where $\vert \tilde{\rho})$ is the state vector in the rotating frame, and $\mathcal{G}_{\text{RWA}} = \sum_k \mathcal{G}^{(k)}_{\text{RWA}}$, with
\begin{align} \label{GRWA_kk} 
\mathcal{G}^{(k)}_{\rm{RWA}} &= i \left[\Delta\right]_{k,k} + i \frac{ \Omega_R}{2} \left( \left[S_x^{\copyright}\right]_{k,k+1} + \left[S_x^{\copyright}\right]_{k,k-1} \right) \nonumber \\
&+ \Gamma_{\rm{tot}}  \left[\mathcal{A}_{\text{SD}}\right]_{k,k}+ \Gamma_{\rm{FD}}\left[ \mathcal{A}_{\rm{FD}}\right]_{k,k}-\frac{1}{2}R_{\rm{op}} \left[ \mathcal{A}_{\rm{SE}}^{(z)}\right]_{k,k} \nonumber \\
&-\Gamma_{\rm{SET}}^{\rm{RbXe}}\langle \mathbf{K} \rangle \cdot  \left[ \boldsymbol{\mathcal{A}}_{\rm{SE}}\right]_{k,k}-\Gamma_{\rm{FE}}\langle \mathbf{K} \rangle \cdot  \left[ \boldsymbol{\mathcal{A}}_{\rm{FE}}\right]_{k,k} \nonumber \\
&-\Gamma_{\text{SE}}^{\rm{RbRb}}\left(   \mathcal{A}_{\text{SE}}^{(k,z)}+\mathcal{A}_{\text{SE}}^{(k, +)} +\mathcal{A}_{\text{SE}}^{(k, -)}\right),
\end{align}
is the effective evolution superoperator. The block matrix  $\left[ \mathcal{X}\right]_{p,q} = \mathcal{P}_p \left[\mathcal{X}\right] \mathcal{P}_q$ describes the coupling between the  $p$-th and $q$-th coherences, with 
\begin{align}
    \mathcal{P}_k = \sum_{\bar{m}} |a,\bar{m})_{k} ~_{k}(a,\bar{m}| + \sum_{\bar{m}} |b,\bar{m})_{-k} ~_{-k}(b,\bar{m}|, \label{Projectionk}
\end{align}
the $k$-th order Zeeman coherence projectors \cite{Feng2025}.

In Eq.~\eqref{GRWA_kk}, we have introduced the transverse SE superoperators
\begin{align}
    \mathcal{A}_{\text{SE}}^{(k, \pm)} &= \tilde{S}_x^{(\mp 1)} \left[\mathcal{A}_{\text{SE}}^{(x)}\right]_{k,k\pm 1} + \tilde{S}_y^{(\mp 1)} \left[\mathcal{A}_{\text{SE}}^{(y)}\right]_{k,k\pm 1} 
\end{align}
and  the longitudinal SE operator
\begin{align}
    \mathcal{A}_{\text{SE}}^{(k, z)} &=  \tilde{S}_z^{(0)} \left[\mathcal{A}_{\text{SE}}^{(z)}\right]_{k,k},
\end{align}
where $\tilde{\mathbf{S}}^{(k)} = (\mathbf{S}\vert \mathcal{P}_{k} \vert \tilde{\rho})=(\mathbf{S} \vert \tilde{\rho}_k)$ is electronic-spin projections on the $k$-th Zeeman coherences in the rotating frame.

The master equation for the first-order Zeeman coherences $\vert \tilde{\rho}_1)$ is obtained by applying the projector $\mathcal{P}_1$ to both sides of Eq.~\eqref{eq:RWA_master_equation}:
\begin{align} \label{FirstOrder1}
\frac{d}{dt} |\tilde{\rho}_1) =& - \left[\mathcal{G}_{\text{RWA}}\right]_{1, 1} \vert\tilde{\rho}_1) - \left[\mathcal{G}_{\text{RWA}}\right]_{1, 0} \vert\tilde{\rho}_0) - \left[\mathcal{G}_{\text{RWA}}\right]_{1, 2} \vert\tilde{\rho}_2),
\end{align}
where the diagonal block is
\begin{align}
\left[\mathcal{G}_{\text{RWA}}\right]_{1, 1} =& i \Delta_1 \mathcal{P}_1 +\Gamma_{\text{tot}} \left[\mathcal{A}_{\text{SD}}\right]_{1,1}
+ \Gamma_{\rm{FD}}\left[ \mathcal{A}_{\rm{FD}} \right]_{1,1} \nonumber \\
-& \frac{R_{\rm{op}}}{2} \left[ \mathcal{A}_{\rm{SE}}^{(z)}\right]_{1,1}- \Gamma_{\rm{SE}}^{\rm{RbRb}}\tilde{S}_{z}^{(0)} \left[ \mathcal{A}_{\rm{SE}}^{(z)}\right]_{1,1} \nonumber\\
-& \Gamma_{\rm{SET}}^{\rm{RbXe}} \langle K_z \rangle \left[ \mathcal{A}_{\rm{SE}}^{(z)}\right]_{1,1}
- \Gamma_{\rm{FE}} \langle K_z \rangle \left[ \mathcal{A}_{\rm{FE}}^{(z)}\right]_{1,1}, \label{GRWA11}
\end{align}
and the off-diagonal blocks are
\begin{align}
\left[\mathcal{G}_{\text{RWA}}\right]_{1,0} =&  \frac{i \Omega_R}{2} \left[S_x^{\copyright}\right]_{1,0} - \Gamma_{\text{SE} }^{\rm{RbRb}} \mathcal{A}_{\text{SE}}^{(1,-)}, \\
\left[\mathcal{G}_{\text{RWA}}\right]_{1,2} =&  \frac{i \Omega_R}{2} \left[S_x^{\copyright}\right]_{1,2} - \Gamma_{\text{SE} }^{\rm{RbRb}} \mathcal{A}_{\text{SE}}^{(1,+)}.\label{Eq:SecondOrderMatrix}
\end{align}
The diagnonal block $\left[\mathcal{G}_{\text{RWA}}\right]_{1, 1}$ couples different first-order Zeeman coherences, while $\left[\mathcal{G}_{\text{RWA}}\right]_{1,0}$ and $\left[\mathcal{G}_{\text{RWA}}\right]_{1,2}$ couples the populations $\vert \tilde{\rho}_0)$ and second-oder Zeeman coherences $\vert \tilde{\rho}_2)$ to $\vert \tilde{\rho}_1)$, respectively. The transverse components of the nuclear-spin expectation value $\langle K_{x,y} \rangle$ do not contribution to the evolution of $\vert \tilde{\rho}_1)$, since $\left[ \mathcal{A}_{\rm{SE}}^{(x,y)}\right]_{1,1}=0$, $\left[ \mathcal{A}_{\rm{FE}}^{(x,y)}\right]_{1,1}=0$, as can be shown through straightforward calculations. The first-order detuning $[\Delta]_{1,1}$ is rewritten as $\Delta_1 \mathcal{P}_1$, with $\Delta_1$ representing the first order frequency detuning.

In the presence of a weak driving field, the system is slightly perturbed from the equilibrium state $|\rho_{\rm{eq}})$, where it is influenced only by optical pumping and spin relaxation. In this case, the population is approximately constant, and $|\tilde{\rho}_0)$ can be approximated by 
\begin{align}
\vert \tilde{\rho}_0) = \mathcal{P}_0 |\rho_{\rm{eq}})\equiv |\bar{\rho}_0).
\end{align}
Since the magnitude of the second order Zeeman coherences $\vert \tilde{\rho}_2)$ is non-negligible only when the Rabi frequency $\Omega_R$ exceeds the characteristic relaxation rates, as indicated by Eq.~\eqref{Eq:SecondOrderMatrix}. we neglect $\vert \tilde{\rho}_2)$ in Eq.~\eqref{Eq:SecondOrderMatrix} under the weak driving field regime. With this weak driving approximation (WDA), the first order coherences $\vert \tilde{\rho}_{1})$ follows the equation
\begin{align}
    \label{FirstOrder1WDA}
    \frac{d}{dt} |\tilde{\rho}_1) &= - \left( \left[\mathcal{G}_{\text{RWA}}\right]_{1, 1} - \Gamma_{\rm SE}^{ \rm{RbRb} } \left[\mathcal{A}_{\text{SE} }^{(\perp)}\right]_{1,1} \right)  \vert \tilde{\rho}_1) -  \Omega_R \vert \bar{\nu}_1 ) \nonumber\\
    & \equiv - \mathcal{G}_{\text{WDA}} \vert \tilde{\rho}_1 ) -  \Omega_R \vert \bar{\nu}_1),
\end{align}
where we have introduced the transverse spin-exchange superoperator 
\begin{align}
    \left[\mathcal{A}_{\text{SE} }^{(\perp)}\right]_{1, 1} &= \left[\mathcal{A}_{\text{SE}}^{(x)}\right]_{1,0} \left[\mathcal{Q}_{x}\right]_{0,1} + \left[\mathcal{A}_{\text{SE}}^{(y)}\right]_{1,0} \left[\mathcal{Q}_{y}\right]_{0,1}, \label{GWDA}
\end{align}
and the matrices $\left[ \mathcal{Q}_{x,y} \right]_{0,1}$ are associated with the equilibrium state $\vert \rho_{\rm{eq}})$ as 
\begin{align}
\left[\mathcal{Q}_{x,y}\right]_{0, 1} = \mathcal{P}_0 \vert \rho_{\text{eq}}) (S_{x, y}\vert \mathcal{P}_1.\label{Qxy}
\end{align}
The inhomogeneous term in Eq.~\eqref{FirstOrder1WDA} arises from the transverse driving field, with the state vector $\vert \bar{v}_1) $ given by 
\begin{align}
\label{SourceTerm}
\vert \bar{v}_1) = \frac{i}{2} \left[ S_{x} ^{\copyright} \right] _{1,0} \vert \bar{\rho}_0).
\end{align}
As indicated by Eqs.~\eqref{FirstOrder1WDA} and \eqref{SourceTerm}, $\vert \bar{v}_1) $ acts as a source term, generating first-order coherences by pumping the populations at a rate $\Omega_R$.

It is pointed out by Appelt \cite{Appelt1998}, the equilibrium state of alkali-metal atoms is well described by the spin temperature distribution
\begin{align}
\rho_{\rm{eq}}= \frac{e^{\beta F_z}}{Z},
\end{align}
when the ensemble is subject to longitudinal pumping and atomic collision, with the spin temperature parameter $\beta$ related to the electron-spin polarization $P$ by 
\begin{align}
P=2 \langle S_z \rangle = \tanh{\frac{\beta}{2}}.
\end{align}

For typical gas components and operating conditions in NMRG applications, the third-body collisional rates $\Gamma_{\rm{FE}}$ and $\Gamma_{\rm{FD}}$ are much smaller than the total binary collisional rate $\Gamma_{\rm{SD}}$, and the equilibrium polarization $P$ is approximately given by \cite{Appelt1998}:
\begin{align}
P = \frac{ R_{\rm{op}} + 2 \langle K_z \rangle \Gamma_{\rm{SET}}^{\rm{RbXe}} }{\Gamma_{\rm{SD}}+ R_{\rm{op}} }.
\end{align}

\subsection{The observable and linewidth}

In typical magnetometer and NMRG applications, the dynamics of the alkali-metal atoms is monitored  using a linearly polarized laser beam perpendicular to the pumping beam. The polarization of the probe light is modified by the transverse spin components $\langle S_{x,y} \rangle$. Suppose that $\langle S_{x} \rangle$ is measured as the observable. The steady-state signal is given by 
\begin{align}
\langle S_x \rangle_{\infty} = (S_x|\tilde{\rho}_{1,\infty}) e^{-i \omega t} + \text{c.c},
\end{align}
where $\vert \tilde{\rho}_{1,\infty}) $ represents the steady-state first-order coherences in the rotating frame.

The value of $\vert \tilde{\rho}_{1,\infty}) $ can be determined from Eq.~\eqref{FirstOrder1WDA}. By expanding the superoperator $\mathcal{G}_{\text{WDA}}$ in terms of its right eigenvectors $\vert \lambda_j )$ and left-eigenvectors $\{ \lambda_j \vert$:
\begin{align}
\mathcal{G}_{\text{WDA}} = \sum_{j} (i\Delta_{1} +\lambda_{j}) \vert \lambda_j) \{\lambda_j \vert,
\end{align}
we get
\begin{align}
\vert \tilde{\rho}_{1,\infty}) = - \sum_{j} \frac{\Omega_R}{ i \Delta_{1} + \lambda_j  } |\lambda_j) \{\lambda_j \vert \bar{v}_1) .
\end{align}
The amplitude of time-dependent observable 
\begin{equation}
    \label{eq:observable}
    (S_x \vert \tilde{\rho}_{1,\infty}) = - \frac{i\Omega_R}{2}\sum_j \frac{w_j}{i\Delta_1 + \lambda_j }.
\end{equation}
is a superposition of several Lorentzian curves, where each term is weighted by
\begin{align}
w_{j} = \{ \lambda_j \left[ S_x^{\copyright}\right]_{1,0} \left[ \mathcal{Q} \right]_{0,1} \vert \lambda_j).
\end{align}
As demonstrated in \cite{Feng2025}, only one dominant weight factor typically contributes significantly, allowing the resonance line shape to be well approximated by a single Lorentzian function, and the linewidth is given by the real part of the smallest eigenvalue of $\mathcal{G}_{\rm{WDA}}$. In the following sections, we will directly diagonalize $\mathcal{G}_{\rm{WDA}}$ to obtain its minimal eigenvalue and, consequently, the linewidth. The deviations from this single minimal-eigenvalue approximation is dissused in  Appendix~\ref{AppendixC}.

\section{Results and analyses}

\subsection{Collision rates}\label{Section_Rates}
\begin{figure}[h]
    \centering
    \includegraphics[width=8.5cm]{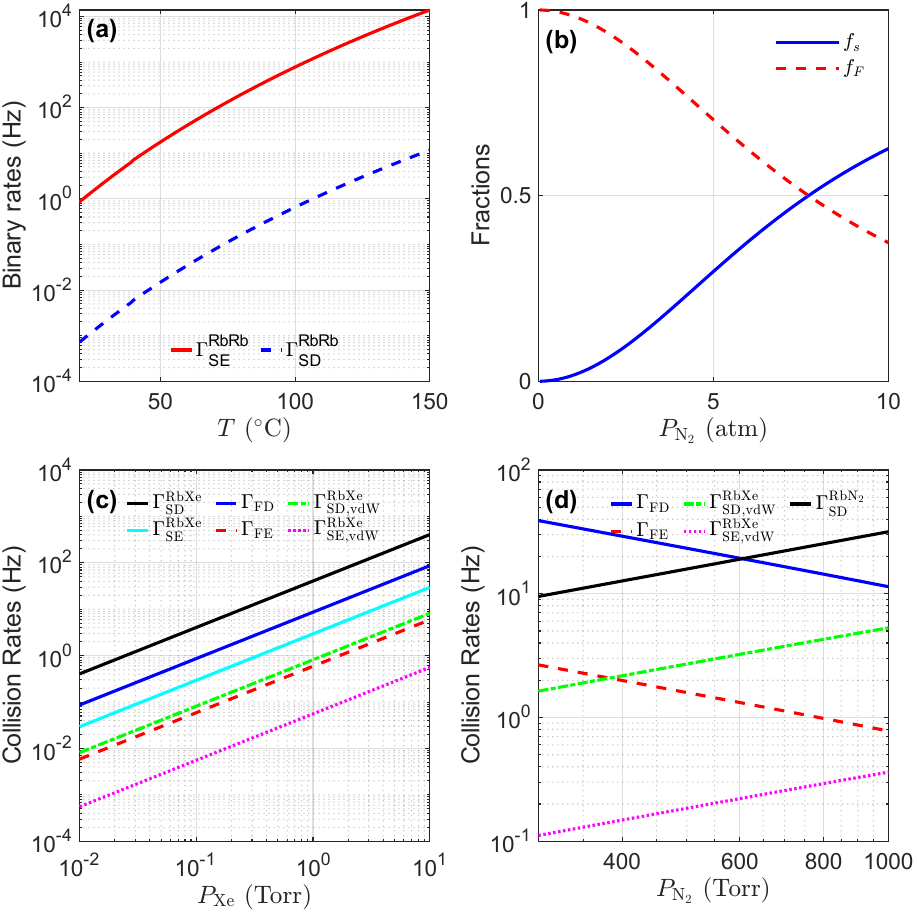}
    \caption{
        Calculated collision rates affecting Rb spin relaxation under NMRG conditions. 
        (a) Temperature dependence of Rb-Rb SD and SE collision rates.
        (b) Dependence of the fractional weights $f_s$ and $f_F$ on nitrogen pressure.
        (c) Xe-density dependence of the Xe-related collsion rates.
        (d)  Dependence of nitrogen-related collision rates on $P_{\mathrm{N}_2}$.
    }
    \label{CollisionRates}
\end{figure}

Figure~\ref{CollisionRates} summarizes the key collisional relaxation rates relevant to Rb spin dynamics under typical NMRG conditions, plotted against several experimental variables. This overview provides a quantitative foundation for analyzing how each process contributes to the total resonance linewidth.

As shown in Fig.~\ref{CollisionRates}(a), the Rb–Rb spin-exchange rate $\Gamma_{\rm{SE}}^{\rm{RbRb}}$ increases steeply with temperature, owing to its strong dependence on atomic vapor density. At typical operating temperatures above $100~^\circ\mathrm{C}$, $\Gamma_{\rm{SE}}^{\rm{RbRb}}$ exceeds all other relevant collisional rates by at least an order of magnitude. This justifies the use of perturbative treatments in linewidth calculations, where other relaxation processes are treated as small corrections to the dominant Rb–Rb SE interaction.

Among the Rb–Xe interaction channels plotted in Fig.~\ref{CollisionRates}(c), the binary spin-destruction rate $\Gamma_{\rm{SD}}^{\rm{RbXe}}$ and the slow vdW-induced F-damping rate $\Gamma_{\rm{FD}}$ are the most significant contributors to Rb relaxation. These two processes dominate the Xe-related effects on the linewidth, while the rates associated with spin exchange and fast vdW-induced damping remain comparatively small.

Figures~\ref{CollisionRates}(b) and \ref{CollisionRates}(d) highlight the dual role of nitrogen in Rb spin relaxation. On one hand, increasing $P_{\rm{N}_2}$ enhances the binary Rb–$\mathrm{N}_2$ spin-destruction rate $\Gamma_{\rm{SD}}^{\rm{RbN}_2}$, directly contributing to S-damping. On the other hand, nitrogen serves as the third-body partner in the formation of Rb–Xe–$\mathrm{N}_2$ van der Waals molecules, which mediate the slow vdW F-damping process. As shown in Fig.~\ref{CollisionRates}(d), $\Gamma_{\rm{SD}}^{\rm{RbN}_2}$ increases linearly with $P_{\rm{N}_2}$, whereas the F-damping rate $\Gamma_{\rm{FD}}$ decreases due to the shorter lifetime of vdW molecules at higher nitrogen pressures. This opposing dependence of the dominant relaxation rates on $P_{\rm{N}_2}$—with $\Gamma_{\rm{SD}}^{\rm{RbN}_2}$ increasing and $\Gamma_{\rm{FD}}$ decreasing—leads to a competition between binary and vdW-mediated relaxation processes. Consequently, an optimal nitrogen pressure may exists at which the total linewidth reaches a minimum.

\subsection{Influence of Xe Polarization and Density on the Rb Linewidth}

\subsubsection{Xe Polarization}

The polarization of xenon atoms contributes to rubidium (Rb) spin relaxation through both binary spin-exchange (SE) collisions and three-body van der Waals (vdW) interactions, characterized by the rates $\Gamma_{\rm{SET}}^{\rm{RbXe}}$ and $\Gamma_{\rm{FE}}$, respectively.

\begin{figure}[h]
    \centering
    \includegraphics[width=8.5cm]{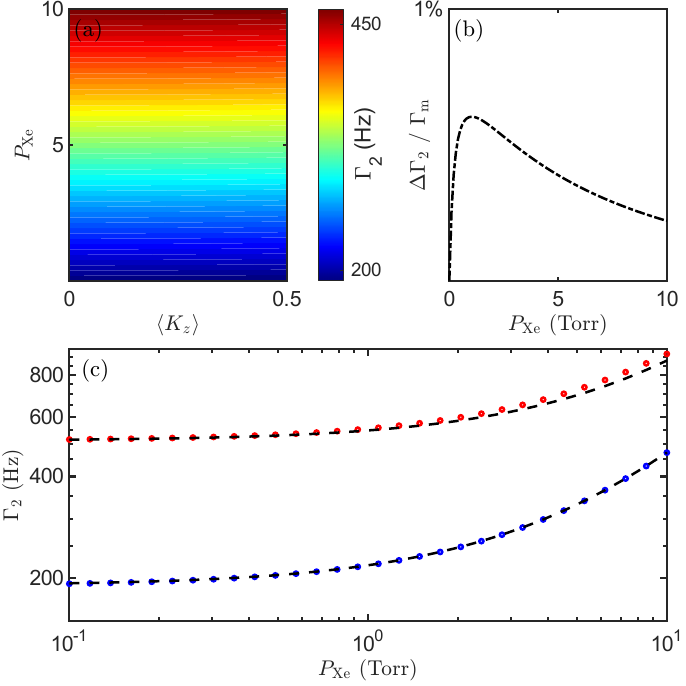}
    \caption{(a) The linewidth $\Gamma_2$ as a function of Xe polarization $\langle K_z \rangle$ and Xe density. (b) The relative linewidth variation $\Delta\Gamma_2 / \Gamma_{\rm{m}} = (\Gamma_{2,\rm{max}} - \Gamma_{2,\rm{min}})/\Gamma_{2,\rm{m}}$ at different Xe densities. Here, $\Gamma_{2,\rm{max}}$ and $\Gamma_{2,\rm{min}}$ are the maximum and minimum linewidths obtained by varying Xe polarization from zero to full polarization ($\langle K_z \rangle = 1/2$), and $\Gamma_{2,\rm{m}}$ denotes the mean linewidth over $\langle K_z \rangle$.
    (c) Dependence of the resonance linewidth $\Gamma_2$ on Xe density. Red and blue dots denote numerically calculated values of $\Gamma_2$ for optical pumping rates $R_{\rm{op}}=2\pi\times 1000~\mathrm{Hz}$ and $R_{\rm{op}}=2\pi\times 1~\mathrm{Hz}$, respectively. The dashed curves are depicted according to Eqs.~\eqref{Gamma2_0} and \eqref{HighPolarization}.  Other parameters used in the calculation are: $T = 110~^{\circ}\mathrm{C}$, $P_{\mathrm{N}_2}=450~\mathrm{Torr}$, and $\langle K_z \rangle = 0$.}
    \label{XePolarizationAndDensity}
\end{figure}

Figure~\ref{XePolarizationAndDensity}(a-b) shows the dependence of the Rb linewidth $\Gamma_2$ on the Xe nuclear polarization $\langle K_z \rangle$. The results indicate that the influence of Xe polarization on the Rb resonance is minor, typically leading to variations at the $\sim1\%$ level. To understand this weak dependence, we analyze the structure of the relevant relaxation superoperators and the magnitudes of the corresponding collision rates.

In the coupled tensor basis \cite{Happer1977},
\begin{align}
T_{LM}(FF') = \sum_{m} \vert F,m\rangle \langle F',m-M \vert (-1)^{m-M-F'} C_{F,m;F',M-m}^{L,M},
\end{align}
the relevant relaxation superoperators are given by:
\begin{align}
\left[ \mathcal{A}_{\rm{SD}} \right]_{1,1} &= \frac{1}{16} \, \mathrm{diag}(16,12,9,7,13,11), \\
\left[ \mathcal{A}_{\rm{FD}} \right]_{1,1} &= \mathrm{diag}(10,6,3,1,3,1), \\
\left[ \mathcal{A}_{\mathrm{SE}}^{(z)} \right]_{1,1} &= \left( \begin{array}{cccc|cc}
0 & \frac{\sqrt{105}}{14} & 0 & 0 & 0 & 0 \\
0 & 0 & \frac{\sqrt{70}}{10} & 0 & 0 & 0 \\
0 & \frac{\sqrt{70}}{70} & 0 & \frac{3\sqrt{105}}{40} & 0 & 0 \\
0 & 0 & \frac{\sqrt{105}}{40} & 0 & 0 & 0 \\ \hline
0 & 0 & 0 & 0 & 0 & -\frac{1}{8} \\
0 & 0 & 0 & 0 & -\frac{3}{8} & 0
\end{array} \right), \label{ASEz} \\
\left[ \mathcal{A}_{\mathrm{FE}}^{(z)} \right]_{1,1} &= \left( \begin{array}{cccc|cc}
0 & \frac{5\sqrt{105}}{7} & 0 & 0 & 0 & 0 \\
-\frac{3\sqrt{105}}{7} & 0 & \frac{32\sqrt{70}}{35} & 0 & 0 & 0 \\
0 & -\frac{16\sqrt{70}}{35} & 0 & \frac{3\sqrt{105}}{5} & 0 & 0 \\
0 & 0 & -\frac{\sqrt{105}}{5} & 0 & 0 & 0 \\ \hline
0 & 0 & 0 & 0 & 0 & 3 \\
0 & 0 & 0 & 0 & -1 & 0
\end{array} \right). \label{AFEz}
\end{align}

The transverse spin-exchange superoperator, which depends on the Rb polarization $P$, takes the form:
\begin{align}
\left[ \mathcal{A}_{\mathrm{SE}}^{(\perp)}(P) \right]_{1,1} &=\frac{1}{16(1+P^2)}\times \nonumber \\
 & \begin{pmatrix}
0 & 0 & 0 & 4 \sqrt{\frac{10}{7}} P^3 & 0 & 0 \\
0 & 0 & 0 & 4 \sqrt{6} P^2 & 0 & 0 \\
0 & 0 & 0 &  \sqrt{\frac{15}{7}} P (7 + P^2) & 0 & 0 \\
0 & 0 & 0 &  (5 + 3 P^2) & 0 & 0 \\
0 & 0 & 0 & 0 & 0 & (P - P^3) \\
0 & 0 & 0 & 0 & 0 &  (1 - P^2) 
\end{pmatrix}.
\end{align}

Xe polarization enters the relaxation generator $\mathcal{G}_{\rm{WDA}}$ through the last two terms in Eq.~\eqref{GRWA11}, associated with the collision rates $\Gamma_{\rm{SET}}^{\rm{RbXe}}$ and $\Gamma_{\rm{FE}}$. These rates correspond to off-diagonal elements in $\mathcal{G}_{\rm{WDA}}$, as shown in Eqs.~\eqref{ASEz} and \eqref{AFEz}. As illustrated in Fig.~\ref{CollisionRates}, these rates are significantly smaller than the dominant contributions $\Gamma_{\rm{tot}}$ and $\Gamma_{\rm{FD}}$, implying that the eigenvalues of $\mathcal{G}_{\rm{WDA}}$ are only weakly perturbed by Xe polarization.

\subsubsection{Xe density}

As discussed in the preceding section, all collision rates involving xenon increase linearly with Xe density, as illustrated in Fig.\ref{CollisionRates}(c). As the Xe density rises, two primary mechanisms contribute to the broadening of the alkali-metal resonance linewidth. First, the total S-damping rate is enhanced due to both binary Rb–Xe spin-destruction collisions and fast vdW processes. Second, the F-damping rate increases via the slow vdW mechanism.

At typical operating temperatures, the Rb–Rb spin-exchange rate $\Gamma_{\rm{SE}}^{\rm{RbRb}}$ dominates over other collision processes. This hierarchy allows the use of perturbation theory to estimate the eigenvalues of the evolution superoperator. For low optical pumping rates $R_{\rm{op}}$, the Rb polarization remains small ($P \ll 1$), such that only the terms proportional to $\Gamma_{\rm{tot}}$ and $\Gamma_{\rm{FD}}$ contribute significantly. Retaining only these terms yields the leading-order approximation for the linewidth:
\begin{align}
\Gamma_2^{(0)} =& \Gamma_{\rm{FD}}+ \frac{7}{16}\Gamma_{\rm{tot}} - \frac{5}{16}\Gamma_{\rm{SE}}^{\rm{RbRb}} \nonumber \\
=&\frac{1}{8}\Gamma_{\rm{SE}}^{\rm{RbRb}} +\frac{7}{16}(R_{\rm{op}}+ \Gamma_{\rm{SD}}^{\rm{RbRb}} +  \Gamma_{\rm{SD}}^{\rm{RbN}_2}  ) \nonumber \\
+& \Gamma_{\rm{FD}} + \frac{7}{16}\left[ \Gamma_{\rm{SET}}^{\rm{RbXe}} + \Gamma_{\rm{SD,vdW}}^{\rm{RbXe}} +  \Gamma_{\mathrm{SD}}^{\rm{RbXe}}    \right].
\label{Gamma2_0}
\end{align} 
Compared to the case where only Rb and $\rm{N}_2$ are present \cite{Feng2025}, the inclusion of Xe introduces additional relaxation mechanisms for Rb, including:
\begin{itemize}
    \item \rm{Rb-Xe SE collisions}: \(\frac{7}{16} \Gamma_{\rm{SET}}^{\rm{RbXe}} \),
    \item \rm{Rb-Xe SD collisions}: \(\frac{7}{16} \Gamma_{\rm{SD}}^{\rm{RbXe}} \),
    \item \rm{S-damping from vdW collisions}: \(\frac{7}{16} \Gamma_{\rm{SD,vdW}}^{\rm{RbXe}}  \).
    \item \rm{F-damping due to vdW collisions}: \( \Gamma_{\rm{FD}} \).
\end{itemize}
As shown in Fig.~\ref{XePolarizationAndDensity}(c), this lowest-order approximation captures the dominant physical effects in the low-polarization regime.


At sufficiently high optical pumping rates, the system enters the high-polarization regime. In this case, the linewidth can be approximated by:
\begin{align}
\Gamma_{2}^{(\mathrm{H})} = \frac{1}{4}R_{\rm{op}} + \frac{1}{2}\Gamma_{\rm{SDP}}+\Gamma_{\rm{FD}} + \frac{5\Gamma_{\rm{SE}}^{\rm{RbRb}}}{24R_{\rm{op}}}
\Gamma_{\rm{SDP}}, \label{HighPolarization}
\end{align}
where
\begin{align}
\Gamma_{\rm{SDP}}=\Gamma_{\rm{SD}}^{\rm{RbRb}} + \Gamma_{\rm{SD}}^{\rm{RbN}_2} + \Gamma_{\rm{SD}}^{\rm{RbXe}} + \Gamma_{\mathrm{SET}}^{\mathrm{RbXe}} + \Gamma_{\rm{SD,vdW}}^{\rm{RbXe}}.
\end{align}
Figure~\ref{XePolarizationAndDensity}(c) shows that Eq.~\eqref{HighPolarization} agrees well with numerical results in the high-polarization regime. The deviations observed at high Xe densities are attributed to reduced Rb polarization with increasing Xe density.

In the intermediate pumping regime, numerical diagonalization of the evolution superoperator is required to accurately compute the linewidth, since the longitudinal spin-exchange term $P\Gamma_{\rm{SE}}^{\rm{RbRb}} \left[\mathcal{A}_{\rm{SE}}^{(z)}\right]_{1,1}/2$, the optical pumping term $R_{\rm{op}}\left[\mathcal{A}_{\rm{SE}}^{(z)}\right]_{1,1}/2$, and the transverse spin-exchange term $\Gamma_{\rm{SE}}^{\rm{RbRb}} \left[ \mathcal{A}_{\mathrm{SE}}^{(\perp)}(P) \right]_{1,1}$ can no longer be treated as small perturbations.

\subsection{The effects of $\rm{N}_2$ on Rb linewidth}

\begin{figure}[h]
    \centering
    \includegraphics[width=8.5 cm]{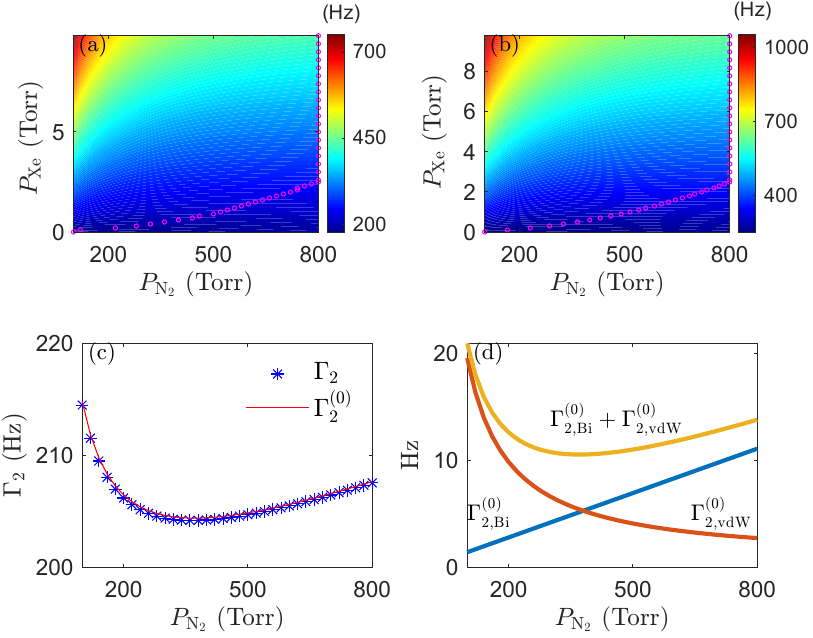}
    \caption{ (a) Dependence of linewidth $\Gamma_2$ on $\rm{N}_2$ and $\rm{Xe}$ densities in the low-polarization regime, with a pumping rate of $R_{\rm{op}}= 2\pi \times 1~\rm{Hz}$. (b) Dependence of linewidth $\Gamma_2$ on $\rm{N}_2$ and Xe densities in the high-polarization regime, with  $R_{\rm{op}}= 2\pi \times 1000~\rm{Hz}$. (c) Dependence of $\Gamma_2$ on $\rm{N}_2$ pressure at fixed Xe pressure $P_{\rm{Xe}}=0.5$ Torr in the low-polarization regime. (d) Contributions of $\Gamma_{2,\rm{Bi}}^{(0)}$, $\Gamma_{2,\rm{vdW}}^{(0)}$, and their sum to the total linewidth as functions of $\rm{N}_2$ pressure [see Eqs.~\eqref{Gamma2_B} and \eqref{Gamma2_vdW}]. The magenta dots in (a) and (b) indicate the critical $\rm{N}_2$ pressure at which $\Gamma_2$ reaches its minimum for a given $P_{\rm{Xe}}$. All calculations assume  $\langle K_z \rangle = 0$ and a temperature of $T =110~^{\circ}\text{C}$   .}
    \label{GammaN2}
\end{figure}
The $\rm{N}_2$ molecules influence the dynamics of the Rb atoms in two distinct ways: through binary spin-destruction collisions with the Rb atoms, and by acting as participants in the formation of Rb-Xe van der Waals molecules. The Rb-$\rm{N}_2$ SD collisions induce S-damping effects, which completely disrupt the average electronic spin. In addition, $\rm{N}_2$ molecules, as third-party participants in the formation of Rb-Xe vdW molecules, affect the vdW formation rate $1/T_{\rm{vdWA}}$ and the lifetimes $\tau$, thereby influencing the relevant collision rates \cite{Zeng1985,Appelt1998}.

The dependence of linewidth $\Gamma_2$ on $\rm{N}_2$ density is illustrated in Fig.~\ref{GammaN2}(a), in the low-polarization regime, $\Gamma_2$ exhibits a non-monotonic dependence on $P_{\rm{N}_2}$, a trend more clearly illustrated in Fig.~\ref{GammaN2}(c). This non-monotonic behavior results from the competition between Rb-$\rm{N}_2$ binary spin-destruction collisions and third-body interactions that influence van der Waals molecule formation. The total linewidth $\Gamma_2^{(0)}$ can be decomposed into three contributions:
\begin{align} 
\Gamma_2^{(0)} = \Gamma_{2,0}^{(0)} + \Gamma_{2,\rm{Bi}}^{(0)} + \Gamma_{2,\rm{vdW}}^{(0)}, 
\end{align}
where
\begin{align} 
\Gamma_{2,0}^{(0)} = \frac{1}{8} \Gamma_{\rm{SE}}^{\rm{RbRb}} + \frac{7}{16} (R_{\rm{op}} + \Gamma_{\rm{SE}}^{\rm{RbXe}} + \Gamma_{\rm{SD}}^{\rm{RbXe}} + \Gamma_{\rm{SD}}^{\rm{RbRb}}) \label{Gamma2_0} 
\end{align}
is independent of $\rm{N}_2$,
\begin{align} 
\Gamma_{2,\rm{Bi}}^{(0)} = \frac{7}{16} \Gamma_{\rm{SD}}^{\rm{RbN}_2} = \frac{7}{16} \left[ \rm{N}_2 \right] \sigma_{\rm{SD}}^{\rm{RbN}_2} \bar{v}_{\rm{RbN}_2} \label{Gamma2_B} 
\end{align}
is proportional to the $\rm{N}_2$ density $\left[ \rm{N}_2 \right]$, and
\begin{align} 
\Gamma_{2,\rm{vdW}}^{(0)} = \Gamma_{\rm{FD}} + \frac{7f_s}{16} \left( \frac{2 \phi_{\gamma}^2}{3T_{\mathrm{vdwA}}} + \frac{\phi_{\alpha}^2}{2T_{\mathrm{vdwA}}} \right) 
\label{Gamma2_vdW} 
\end{align}
is approximately inversely proportional to $\left[ \rm{N}_2 \right]$ \cite{Zeng1985,Song2021}, as demonstrated in Fig.~\ref{GammaN2}(d). The opposing dependencies of $\Gamma_{2,\rm{Bi}}^{(0)}$ and $\Gamma_{2,\rm{vdW}}^{(0)}$ suggest the existence of an optimal Xe density, $P_{\rm{Xe}}$, at which $\Gamma_2^{(0)}$ reaches its minimum, as shown in Fig.~\ref{GammaN2}(c).

The dependence of the linewidth $\Gamma_2$ on $\rm{N}_2$ density in the high-polarization regime is calculated and presented in Fig.~\ref{GammaN2}(b). Similar to the low-polarization case, $\Gamma_2$ exhibits a non-monotonic dependence on $\rm{N}_2$ density. However, in the high-polarization regime, the calculation of $\Gamma_2$ requires the direct diagonalization of the relaxation superoperator $\mathcal{G}_{\rm{WDA}}$.

\subsection{Influence of Xe on the light-narrowing effects}

Light-narrowing effects were first observed by Appelt et al.~\cite{Appelt1999} in high-pressure optical-pumping cells operating in the high-field regime, where Zeeman sublevels within each hyperfine multiplet are well resolved. Their study demonstrated that increasing optical pumping intensity led to a pronounced reduction in the magnetic resonance linewidth. In the present work, we explore light-narrowing behavior in the Zeeman-unresolved regime~\cite{Feng2025}, and investigate how xenon affects the conditions under which this phenomenon occurs.

\begin{figure}[h]
    \centering
    \includegraphics[width=8.5 cm]{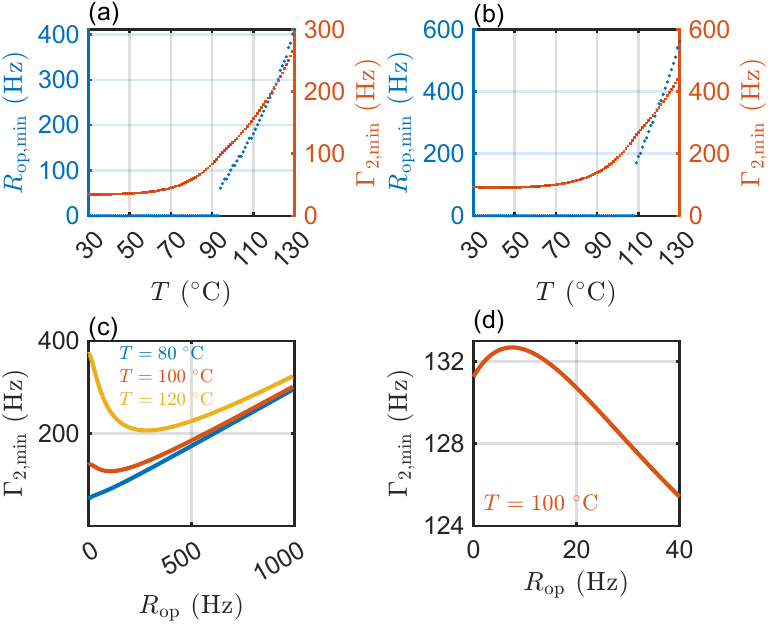}
    \caption{ (a) The minimal linewidth $\Gamma_{2,\rm{min}}$ as a function of cell temperature. The blue dots indicate the values of the optical pumping rate $R_{\rm{op}}$ at which $\Gamma_2$ reaches its minimum. The calculations are performed with $P_{\rm{N}_2} = 450~\rm{Torr}$, $\langle K_z \rangle = 0$, and $P_{\rm{Xe}} = 1~\rm{Torr}$.
(b) Same as (a), but with an increased Xe density of $P_{\rm{Xe}} = 3~\rm{Torr}$.
(c) The linewidth $\Gamma_2$ as a function of the pumping rate $R_{\rm{op}}$ at different cell temperatures. (d) A zoomed-in view of (c) for $T = 100~^\circ\rm{C}$, showing the dependence of $\Gamma_2$ on $R_{\rm{op}}$ in the range of 0–40 Hz.}
    \label{Lightnarrowing}
\end{figure}

Figure~\ref{Lightnarrowing}(a) and (b) show the minimal linewidth $\Gamma_{2,\rm{min}}$ as a function of cell temperature at different Xe pressures. Two trends are evident: (i) light narrowing only occurs above a characteristic temperature threshold $T_{\rm{c}}$, and (ii) both $T_{\rm{c}}$ and $\Gamma_{2,\rm{min}}$ increase with Xe density. These results suggest that Xe plays a critical role in determining the onset and depth of the light-narrowing effect.

To understand this behavior, we analyze the linewidth evolution with pumping rate $R_{\rm{op}}$, as shown in Fig.~\ref{Lightnarrowing}(c) and its zoomed-in view in (d). For $T < T_{\rm{c}}$, $\Gamma_2$ increases monotonically with $R_{\rm{op}}$, indicating the absence of light narrowing. In contrast, when $T > T_{\rm{c}}$, $\Gamma_2$ first increases, reaches a maximum in the low-polarization regime, and then decreases with increasing $R_{\rm{op}}$, exhibiting the characteristic light-narrowing profile.

This transition is governed by the balance between the Rb–Rb spin-exchange rate $\Gamma_{\rm{SE}}^{\rm{RbRb}}$ and the total S-damping rate $\Gamma_{\rm{SDP}}$. As discussed in Ref.~\cite{Feng2025}, the curvature of $\Gamma_2$ versus $R_{\rm{op}}$ in the low-polarization regime (i.e., Fig.~\ref{Lightnarrowing}(d)) flips sign when the ratio $\Gamma_{\rm{SE}}^{\rm{RbRb}}/\Gamma_{\rm{SDP}}$ exceeds a critical threshold. Increasing Xe density enhances the binary SD rate $\Gamma_{\rm{SD}}^{\rm{RbXe}}$, thus increasing $\Gamma_{\rm{SDP}}$. Consequently, a higher cell temperature is required to raise $\Gamma_{\rm{SE}}^{\rm{RbRb}}$ sufficiently to surpass this threshold, leading to an upward shift in $T_{\rm{c}}$.

\section{Conclusions and comments}

In this work, we studied the magnetic resonance linewidth of alkali-metal atoms in the gas cell of a nuclear magnetic resonance gyroscope, with a focus on the role of xenon atoms. Using a density-matrix formalism based on the master equation under the rotating wave approximation and the weak driving limit, we quantitatively analyzed how various collisional mechanisms contribute to the Rb linewidth. Our results reveal that increasing Xe concentration leads to a broader Rb resonance, primarily through two mechanisms: binary spin-destruction collisions and vdW-induced F-damping processes. In contrast, the influence of Xe nuclear polarization $\langle K_z \rangle$ on the Rb linewidth is minor, typically contributing only at the $\sim 1\%$ level.

We also investigated the role of nitrogen and found two competing effects. On one hand, binary Rb–$\mathrm{N}_2$ collisions contribute to spin relaxation and hence broaden the linewidth. On the other hand, the lifetime of 
Rb–Xe–$\mathrm{N}_2$ vdW complexes, which mediate F-damping, is inversely proportional to $\mathrm{N}_2$ pressure, making $\Gamma_{\mathrm{FD}}$ decrease as $\mathrm{N}_2$ density increases. Therefore, $\mathrm{N}_2$ exerts a dual influence on the linewidth, and an optimal buffer gas pressure exists that minimizes the overall Rb relaxation rate.

In addition, we examined how Xe affects the light-narrowing phenomenon. Our analysis shows that light narrowing only emerges when the cell temperature exceeds a characteristic threshold $T_{\rm c}$, which itself increases with Xe concentration. Moreover, the minimum achievable linewidth $\Gamma_{2,\rm{min}}$ becomes larger as Xe density increases, indicating that higher Xe levels hinder the effectiveness of light narrowing.

For practical NMRG implementations, both $^{129}\mathrm{Xe}$ and $^{131}\mathrm{Xe}$ isotopes are commonly used to suppress magnetic bias drift \cite{Walker2016, Eklund2008}. In this study, we focused on $^{129}\mathrm{Xe}$ due to the lack of reliable cross-section data for $^{131}\mathrm{Xe}$. Future measurements of $^{131}\mathrm{Xe}$ collisional parameters would allow more accurate modeling. The theoretical framework developed here can be readily extended to cases involving mixed Xe isotopes, providing a more complete understanding of spin relaxation and resonance linewidth behavior in alkali–noble gas systems used in NMRG applications.

\section{Acknowledgement}
This work is supported by XXXX.

\appendix
\section{Binary collision processes} 
In this section, we provide a detailed introduction to the calculation of collision rates, which are essential for determining the magnetic resonance linewidth. The dynamics of Rb atoms involve two distinct collisional processes: binary collisions and van der Waals molecular interactions. The key parameters used to describe these processes are summarized in Table~\ref{Parameters}.

The binary collision rates are determined by the collisional cross-section, the average relative velocity, and the density of the colliding species:
\begin{align}
\Gamma_{\mathrm{SD},\mathrm{SE}}^{\mathrm{RbX}} = [X] \sigma_{\mathrm{SD},\mathrm{SE}}^{\mathrm{RbX}} \bar{v}_{\mathrm{RbX}},
\end{align}
where $[X]$  represents the number density of atom $\mathrm{X}$, which collides with Rb atoms, and
\begin{align}
 \bar{v}_{\mathrm{RbX}} = \sqrt{\frac{8 k_{B} T}{\pi \mu}},
 \end{align}
denotes the average relative velocity between a Rb atom and a $\rm{X}$ atom, with $\mu$ being the reduced mass of the colliding atoms.

The number density of Rb atoms can be estimated using the empirical formula from Ref.~\cite{Alcock1984}:
\begin{align}
[\mathrm{Rb}] = \frac{1}{T} \begin{cases}
10^{21.866 + 4.857 - \frac{4215}{T}}, & T < (39.3 + 273.15)\rm{K} \\
10^{21.866 + 4.312 - \frac{4040}{T}}, & T > (39.3 + 273.15)\rm{K}
\end{cases},
\end{align}
which provides the Rb density in $\mathrm{cm}^{-3}$.

The density of $\mathrm{N}_2$ and $^{129}\mathrm{Xe}$ can be estimated by the ideal gas law
\begin{align}
[X] = \frac{p}{k_{B}T_0},
\end{align}
where $p$ is the gas pressure and $T_0$ is the temperature at which the gas was initially filled \cite{Zeng1985}.
\begin{table}[btp]
\centering
\renewcommand{\arraystretch}{1.3}
\begin{tabular}{|c|c|c|}
        \toprule
        Parameters & Values & Refs.  \\
        \hline
        $\sigma_{\mathrm{SD}}^{\mathrm{RbN2}}$ & $1 \times 10^{-{22}}~\mathrm{cm}^{-2}$   & \cite{Allred2002}\\
        $\sigma_{\mathrm{SD}}^{\mathrm{RbRb}}$ & $1.6 \times 10^{-17}~\mathrm{cm}^{-2}$   & \cite{Knize1989}\\
        $\sigma_{\mathrm{SD}}^{\mathrm{RbXe129}}$ & $2 \times 10^{-19}~\mathrm{cm}^{-2}$      & \cite{Walker1989}\\
        $\sigma_{\mathrm{SE}}^{\mathrm{RbRb}}$  & $1.9 \times 10^{-{14}}~\mathrm{cm}^{-2}$  &\cite{Ressler1969}\\
        $\sigma_{\mathrm{SE}}^{\mathrm{RbXe129}}$  & $1.6 \times 10^{-{20}}~\mathrm{cm}^{-2}$  &\cite{Walker1989}\\
        $Z$   & $5.0\times 10^{-32}~\mathrm{cm}^6\cdot \mathrm{s}^{-1}$                       & \cite{Zeng1985}\\
        $\gamma N/2\pi$   & $120~\mathrm{MHz}$                                                 & \cite{Walker1989}\\
        $x$               & $3.2$                                                               & \cite{Zeng1985}\\
        $p_0$             & $103~\mathrm{Torr}$                                               & \cite{Zeng1985}\\
        \hline
\end{tabular}
\caption{Basic collision and van der Waals molecular parameters used for calculations.}
\label{Parameters}
\end{table}


\section{The van der Waals molecular process}\label{vdWProcess}

The spin-exchange pumping mechanism between noble-gas atoms and alkali-metal atoms has been extensively studied by Happer and his collaborators \cite{Walker1989,Zeng1985,Happer1984,Appelt1998}. Their research highlighted the crucial role of van der Waals molecular processes in facilitating the transfer of polarization from alkali-metal atoms to noble-gas atoms.

The formation and dissociation of vdW molecules can be described by the reaction \cite{Appelt1998}:
\begin{align} \mathrm{Rb} +\mathrm{Xe} + \mathrm{N}_2 \leftrightarrow \mathrm{RbXe} +\mathrm{N}_2. \end{align}
The key parameters governing these vdW processes are listed in Table \ref{Parameters}. Here, $Z$ represents the rate coefficient for vdW molecule formation, while $\gamma N$ denotes the average spin-rotation interaction strength \cite{Walker1989}. The dimensionless parameter
\begin{align} x = \frac{\gamma N}{\alpha} \end{align}
characterizes the relative strength of the spin-rotation interaction, $\gamma \mathbf{N}\cdot \mathbf{S}$, compared to the spin-exchange interaction, $\alpha \mathbf{K} \cdot \mathbf{S}$ \cite{Zeng1985}.

The spin-rotation interaction contributes to relaxation according to the equation
\begin{align} \frac{d}{dt}\rho = \frac{2 \phi_{\gamma}^2}{3T_{\mathrm{vdwA}}}\left[ f_{s} (\varphi-\rho) + \frac{f_{F}} {[I]^2}(\mathbf{F}\cdot \rho \mathbf{F}-\frac{1}{2}\{{\mathbf{F}\cdot\mathbf{F},\rho}\}) \right], \end{align}
while the nuclear-electron spin-exchange interaction leads to the evolution of $\rho$ as
\begin{align} 
\frac{d}{dt}\rho =& \frac{\phi_{\alpha}^2}{2T_{\mathrm{vdwA}}}\bigg[ f_{s} \left(\varphi(1+4 \langle \mathbf{K} \rangle \cdot \mathbf{S})-\rho \right)+ \frac{f_F}{[I]^2}  \nonumber\\
\times& \big[ \mathbf{F}\cdot \rho \mathbf{F}-\frac{1}{2}\{{\mathbf{F}\cdot\mathbf{F},\rho}\}
+(\{\mathbf{F},\rho \}-2i \mathbf{F}\times \rho \mathbf{F})\cdot \langle \mathbf{K} \rangle \big] \bigg],
\end{align}
where
\begin{align} \phi_{\alpha}= \frac{\alpha \tau}{\hbar}, \
\phi_{\gamma} = \frac{\gamma N \tau}{\hbar} \label{Phig} \end{align}
represent the average phase angles induced by spin-exchange and spin-rotation interactions, respectively. The value of $\phi_{\gamma}$ is determined by the nitrogen gas pressure inside the cell \cite{Zeng1985},
\begin{align} \phi_{\gamma}= \frac{p_{0}}{p_{\mathrm{N}_2}}, \end{align}
where the characteristic pressure $p_0$ is given in Table \ref{Parameters}.
The fractions
\begin{align} 
f_s = \frac{1}{1+(\omega_{\mathrm{hf}}\tau)^2}, \
f_F = \frac{(\omega_{\mathrm{hf}}\tau)^2}{1+(\omega_{\mathrm{hf}}\tau)^2}, \end{align}
describe the proportion of vdW molecules experiencing “very short” and “short” lifetimes, respectively \cite{Appelt1998}. These fractions depend on the hyperfine splitting frequency $\omega_{ \mathrm{hf} }$ and the vdW molecule lifetime $\tau$, which can be determined using Eq.~\eqref{Phig}.

The vdW molecule formation rate per Rb atom, $1/T_{\mathrm{vdwA}}$, is given by
\begin{align} \frac{1}{T_{\mathrm{vdwA}}} = Z [\mathrm{N}_2][\mathrm{Xe}]. \end{align}
Here, the third-body collision partner is assumed to be nitrogen ($\mathrm{N}_2$), as its density significantly dominates under typical experimental conditions.


\section{Beyond the single minimal-eigenvalue approximation}\label{AppendixC}
\begin{figure}[h]
    \centering
    \includegraphics[width=8.5 cm]{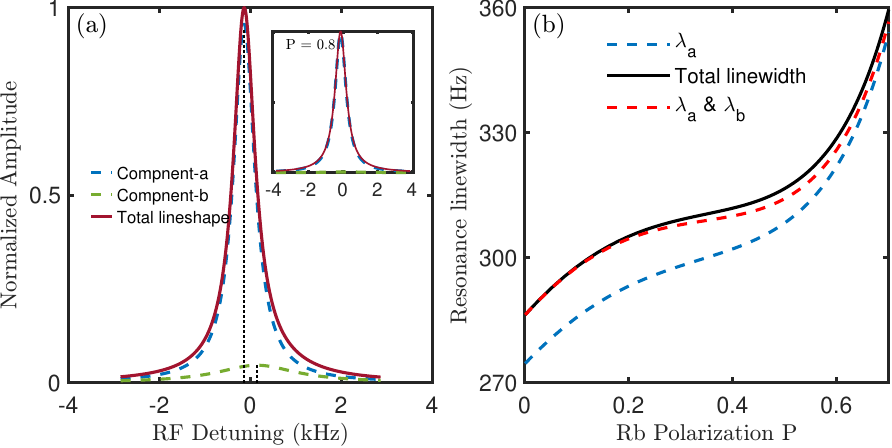}
    \caption{(a) Calculated lineshapes with numerical method, single minimal-eigenvalue approximation and two-eigenvalues approximation. The red solid curve represents the lineshape calculated by numerically solving the master equation \eqref{MasterEq_H}. The blue dashed curve represents the lineshape calculated by single minimal-eigenvalue approximation. The greed dashed curve represents the contribution from $\lambda_b$. The gap between  the two vertical gray lines denoted the resonance-frequency difference in $a$ and $b$ subspace. (b) Variations of linewidth with polarization. The black solid curve represents numerically calculated results, the blue dashed represents the results of single minimal-eigenvalue approximation, and red dashed curve denotes the results of two-eigenvalues approximation. Parameters used in these graphs are: $P_{\rm{N}_2}=450$ Torr, $P_{\rm{Xe}}=3$ Torr, $\langle K_z\rangle =0$, and $T=110~^{\circ} \mathrm{C} $.    }
    \label{FigS2}
\end{figure}
In the main text, the linewidth $\Gamma_2$ of the alkali-metal magnetic resonance was approximated by calculating the minimal eigenvalue of the weak-driving approximation superoperator $\mathcal{G}_{\rm{WDA}}$. This approach provides a simple and effective method for analyzing the dependence of the linewidth on various experimental parameters, such as xenon density, polarization, optical pumping rate, and temperature.

However, it is important to recognize that the single minimal-eigenvalue approximation represents only the leading-order treatment. The contributions from the second-smallest eigenvalue of 
$\mathcal{G}_{\rm{WDA}}$ are less negligible  in the low-polarization regime. 

Figure~\ref{FigS2}(a) compares the lineshapes calculated using three methods: numerical solution of the full master equation, single minimal-eigenvalue approximation, and a two-eigenvalue approximation. Here, the $a$-component corresponds to the lineshape associated with the minimal eigenvalue $\lambda_a$, as adopted in the main text, while the $b$-component represents the contribution from the second-smallest eigenvalue $\lambda_b$. These additional contributions from $\lambda_b$ slightly broaden the overall resonance profile and introduce deviations from a purely Lorentzian shape. As the polarization $P$  increases, the influence of $\lambda_b$ diminishes, and the single minimal-eigenvalue approximation becomes increasingly accurate.

Figure~\ref{FigS2}(b) illustrates the differences between the single minimal-eigenvalue approximation and the two-eigenvalue approximation at various polarization levels. The two-eigenvalue approximation provides a more accurate prediction of the linewidth across the range of polarizations. The remaining discrepancy between the numerically calculated total linewidth and the two-eigenvalue approximation arises from deviations of the steady-state distribution from the ideal spin-temperature distribution, with such deviations becoming more significant at high polarization.
Nevertheless, as shown in Fig.~\ref{FigS2}(b), the single minimal-eigenvalue approximation still yields a reasonably accurate linewidth, with errors typically $\leq5\%$, while the two-eigenvalue approximation offers improved accuracy.


\begin{thebibliography}{99}%
\bibitem{Donley2010} ] E. A. Donley, Nuclear magnetic resonance gyroscopes, in \textit{2010 IEEE Sensors} (IEEE, 2010).
\bibitem{Kitching2011} J. Kitching, S. Knappe, and E. A. Donley, IEEE Sens. J. \textbf{11}, 1749 (2011).
\bibitem{Fang2012}J. Fang and J. Qin, Advances in atomic gyroscopes: A view from inertial navigation applications, Sensors \textbf{12}, 6331 (2012).
\bibitem{Meyer2014} D. Meyer and M. Larsen, Gyroscopy and Navigation \textbf{5}, 75 (2014).
\bibitem{Donley2013} E. Donley and J. Kitching, Nuclear magnetic resonance gyroscopes, \textit{Optical Magnetometry} (Cambridge University Press, Cambridge, England, 2013), p. 369.


\bibitem{Walker2016} T. Walker and M. Larsen, Adv. Atom. Molec. Opt. Phys., \textbf{65}, 373 (2016).
\bibitem{Eklund2008}E. J. Eklund, \textit{Microgyroscope Based on Spin-Polarized Nuclei}, Ph.D. thesis, University of California, Irvine (2008).
\bibitem{Gao2024} G. Gao, J. Hu, F. Tang, W. Liu, X. Zhang, B. Wang, D. Deng, M. Zhu and N. Zhao, Phys. Rev. Applied \textbf{21}, 014042 (2024).
\bibitem{Budker2007} D. Budker and M. Romalis, Nat. Phys. \textbf{3}, 227 (2007).
\bibitem{Budker2013} D. Budker and D. F. J. Kimball, \textit{Optical Magnetometry} (Cambridge University Press, Cambridge, 2013).


\bibitem{Happer1977}W. Happer and A. C. Tam, Phys. Rev. A \textbf{16}, 1877 (1977).
\bibitem{Appelt1998}S. Appelt, A. B.-A. Baranga, C. J. Erickson, M. V. Romalis, A. R. Young, and W. Happer, Phys. Rev. A \textbf{58}, 1412 (1998).
\bibitem{Seltzer2008}S. J. Seltzer, Developments in alkali-metal atomic magnetometry, Ph.D. thesis, Princeton University, 2008.
\bibitem{Bouchiat1972} M. Bouchiat, J. Brossel, and L. Pottier, J. Chem. Phys. \textbf{56}, 3703 (1972).
\bibitem{Zeng1985}X. Zeng, Z. Wu, T. Call, E. Miron, D. Schreiber, and W. Happer, Phys. Rev. A \textbf{31}, 260 (1985). 
\bibitem{Happer1984}W. Happer, E. Miron, S. Schaefer, D. Schreiber, W. A. van Wijngaarden, and X. Zeng, Phys. Rev. A \textbf{29}, 3092 (1984).
\bibitem{Happer2010}W. Happer, Y. Jau, and T. Walker, \textit{Optically Pumped Atoms} (Wiley, New York, 2010).
\bibitem{Happer1972}W. Happer, Rev. Mod. Phys. \textbf{44}, 169 (1972).


\bibitem{Feng2025}F. Tang and N. Zhao, Phys. Rev. A \textbf{111}, 013103 (2025).
\bibitem{Ernst1990}R. R. Ernst, G. Bodenhausen, and A. Wokaun, \textit{Principles of Nuclear Magnetic Resonance in One and Two Dimensions} (Clarendon Press, Oxford, 1987).
\bibitem{Song2021} B. Song, Y. Wang, and N. Zhao, Physical Review A \textbf{104}, 023105 (2021).
\bibitem{Appelt1999}S. Appelt, A. Ben-Amar Baranga, A. R. Young, and W. Happer, Phys. Rev. A \textbf{59}, 2078 (1999).


\bibitem{Allred2002}J. C. Allred, R. N. Lyman, T. W. Kornack, and M. V. Romalis, Phys. Rev. Lett. \textbf{89}, 130801 (2002).
\bibitem{Knize1989} R. J. Knize, Phys. Rev. A \textbf{40}, 6219 (1989). 
\bibitem{Walker1989}T. G. Walker, Phys. Rev. A \textbf{40}, 4959 (1989). 
\bibitem{Ressler1969}N. W. Ressler, R. H. Sands and T. E. Stark, Phys. Rev. \textbf{184}, 102 (1969).

\bibitem{Alcock1984}C. B. Alcock, V. P. Itkin, and M. K. Horrigan, Vapor pressure equations for the metallic elements: 298–2500 K, Can. Metall. Q. \textbf{23}, 309 (1984).









\end{thebibliography}
\end{document}